\documentstyle[prl,eqsecnum,aps]{revtex}
\begin{document}
\author{Jian Qi Shen $^{1,}$$^{2}$ \footnote{E-mail address: jqshen@coer.zju.edu.cn, jqshencn@yahoo.com.cn}}
\address{$^{1}$  Centre for Optical
and Electromagnetic Research, Joint Research Centre of Photonics
of The Royal Institute of Technology (Sweden) and Zhejiang
University, Zhejiang University,
Hangzhou Yuquan 310027, P.R. China\\
$^{2}$ Zhejiang Institute of Modern Physics and Department of
Physics, Zhejiang University, Hangzhou 310027, P.R. China}
\date{\today}
\title{The resemblances in mathematical structures between the optical constants
of artificial electromagnetic media and some physical phenomena in
field theory}

\maketitle

\begin{abstract}
This paper demonstrates that there is much similarity in the
mathematical formalisms between the optical constants of
artificial electromagnetic media (such as chiral media,
left-handed media, photonic crystals and EIT media) and some
physical phenomena in field theory, including general relativity,
quantum mechanics, energy band theory, {\it etc.}. The
significance of such comparisons lies in that: (i) the unification
in mathematical descriptions shows that many physical phenomena
and effects, which seem to have no connections between them,
actually share almost the same mathematical structures; (ii) it
can provide clue to us on suggesting more new effects which is
similar in mathematical descriptions to the familiar phenomena in
other areas.

\end{abstract}
\section{Introduction}
This paper demonstrates that there is much similarity between the
mathematical structures of optical constants of artificial
electromagnetic media (such as chiral media, left-handed media,
photonic crystals and EIT media) and some physical phenomena in
field theory, including general relativity, quantum mechanics,
energy band theory, {\it etc.}. The aim of this paper is to show
that the unification in mathematical descriptions can clue us to
the discovery of some new physical effects, by analogy with those
in other fields.

We consider four strong resemblances in the mathematical
structures between the artificial electromagnetic media and the
physical phenomena in field theory:

(i) the generalization of the constitutive relation of regular
media to that of chiral media resembles the extension of the flat
Minkowski metric to the curved-spacetime metric of Riemann
spacetime. In other words, the off-diagonal terms in the
mathematical formalism of the electromagnetic energy density in
chiral media arises from such a generalization;

(iii) since the stationary Maxwellian equation of the time
harmonic wave strongly resembles the stationary Schr\"{o}dinger
equation of matter waves, if the optical refractive index of media
is periodic in space, then such media will give rise to the
so-called photonic band gap structure. This kind of materials is
called the photonic crystals;

(iii) the negative solutions of any quadratic-form relations such
as $E^{2}=p^{2}c^{2}+m_{0}^{2}c^{4}$ and $n^{2}=\epsilon\mu$ will
also have physical meanings and therefore should not be discarded.
For example, for the former relation
$E^{2}=p^{2}c^{2}+m_{0}^{2}c^{4}$, the negative energy solutions
indicate the existence of the antiparticle of electron, and for
the latter $n^{2}=\epsilon\mu$, the negative solutions corresponds
to a new kind of artificial composite materials, which possess
negative refractive indices. This kind of materials is referred to
as the left-handed media or negative refractive index media;

(iv) the similarity between EIT and superconductivity is
interesting. We show that the effective Hamiltonian (describing
the effective interaction between the two ground states of the
three-level system) and the BCS Hamiltonian have features that are
almost the same. In addition, we take into account the destructive
interference in both Cabbibo's theory (1963) and GIM mechanism
(1970) of the Weak Interaction. We state that there exists the
similar destructive interferences in CPT and EIT. The mathematical
formalisms of these destructive interferences in Weak Interaction
and atomic ensembles are almost alike.

The significance of the above four comparisons between the current
typical subjects in materials science and those in field theory
lies in that: (i) the unification in mathematical descriptions
shows that many physical phenomena and effects, which seem to have
no connections between them, actually share almost the same
mathematical structures; (ii) it can provide clue to us on
suggesting more new effects which is similar in mathematical
descriptions to the familiar phenomena in other areas.
\section{Chiral media}
According to the Maxwellian electrodynamics, the energy density of
electromagnetic fields in electromagnetic media is written
$W=(1/2)\left({\bf E}\cdot{\bf D}+{\bf H}\cdot{\bf B}\right)$. For
a linear medium, the relation between the electric displacement
vector ${\bf D}$ and the electric field strength ${\bf E}$ is
${\bf D}=\epsilon\epsilon_{0}{\bf E}$, and ${\bf H}={\bf
B}/\mu\mu_{0}$ between the magnetic induction ${\bf H}$ and the
magnetic field strength ${\bf B}$. Thus the energy density of
electromagnetic fields in this medium can be rewritten as the
following matrix form
\begin{equation}
W=\frac{1}{2}\left( {\begin{array}{*{20}c}
   {{\bf E}}  & {{\bf B}}
\end{array}} \right) \left( {\begin{array}{*{20}c}
   {\epsilon\epsilon_{0}} & {0}  \\
   {0} & {\frac{1}{\mu\mu_{0}}}  \\
\end{array}} \right)\left( {\begin{array}{*{20}c}
   {\cdot{\bf E}}  \\
   {\cdot{\bf B}}  \\
\end{array}} \right)
\label{EQ1}.
\end{equation}
In the chiral media, however, the electromagnetic properties can
be characterized by another constitutive relations. We can select
\begin{equation}
{\bf D}=\epsilon\epsilon_{0}{\bf E}+i\zeta {\bf B},  \quad   {\bf
H}=i\zeta{\bf E}+\frac{{\bf B}}{\mu\mu_{0}}.
\end{equation}
So, the energy density of electromagnetic fields in such media is
\begin{equation}
W=\frac{1}{2}\left( {\begin{array}{*{20}c}
   {{\bf E}}  & {{\bf B}}
\end{array}} \right) \left( {\begin{array}{*{20}c}
   {\epsilon\epsilon_{0}} & {i\zeta}  \\
   {i\zeta} & {\frac{1}{\mu\mu_{0}}}  \\
\end{array}} \right)\left( {\begin{array}{*{20}c}
   {\cdot{\bf E}}  \\
   {\cdot{\bf B}}  \\
\end{array}} \right)
\label{EQ2}.
\end{equation}
This means that the electromagnetic-parameter matrix of the chiral
media can be considered as the off-diagonal generalization of the
diagonal electromagnetic-parameter matrix of the regular linear
materials, that is, we have
\begin{equation}
\left( {\begin{array}{*{20}c}
   {\epsilon\epsilon_{0}} & {i\zeta}  \\
   {i\zeta} & {\frac{1}{\mu\mu_{0}}}  \\
\end{array}} \right)     \Longleftarrow \left( {\begin{array}{*{20}c}
   {\epsilon\epsilon_{0}} & {0}  \\
   {0} & {\frac{1}{\mu\mu_{0}}}  \\
\end{array}} \right).  \label{chiral}
\end{equation}

It should be noted that the above fact is of much interest, since
it is similar to the curved-spacetime generalization for the
metric from the flat spacetime: specifically, in the two
dimensional Minkowski spacetime, the line element squared can be
given as
\begin{equation}
{\rm d}s^{2}=\left( {\begin{array}{*{20}c}
   {{\rm d}x^{0}}  & {{\rm d}x^{1}}
\end{array}} \right) \left( {\begin{array}{*{20}c}
{+1} & {0}  \\
   {0} & {-1}  \\
\end{array}} \right)\left( {\begin{array}{*{20}c}
   {{\rm d}x^{0}}  \\
   {{\rm d}x^{1}}  \\
\end{array}} \right)
\label{EQ3}.
\end{equation}
Note that here the metric matrix is diagonal. However, in the
Riemann spacetime, the metric matrix has the off-diagonal elements
$g_{10}$ and $g_{01}$, and in consequence, the line element
squared is given
\begin{equation}
{\rm d}s^{2}=\left( {\begin{array}{*{20}c}
   {{\rm d}x^{0}}  & {{\rm d}x^{1}}
\end{array}} \right) \left( {\begin{array}{*{20}c}
   {g_{00}} & {g_{10}}  \\
   {g_{01}} & {g_{11}}  \\
\end{array}} \right)\left( {\begin{array}{*{20}c}
   {{\rm d}x^{0}}  \\
   {{\rm d}x^{1}}  \\
\end{array}} \right)
\label{EQ4}.
\end{equation}
Apparently, there exists a nontrivial extension of metric from the
flat Minkowski spacetime to the Riemann spacetime, {\it i.e.},
\begin{equation}
 \left( {\begin{array}{*{20}c}
{+1} & {0}  \\
   {0} & {-1}  \\
\end{array}} \right)  \Longrightarrow    \left( {\begin{array}{*{20}c}
   {g_{00}} & {g_{10}}  \\
   {g_{01}} & {g_{11}}  \\
\end{array}} \right).   \label{relativity}
\end{equation}
This generalization is just the development of special relativity
(1905) to the general relativity (1916). We think there is some
interesting similarity between (\ref{chiral}) and
(\ref{relativity}), at least in the research fashion of looking
for potentially new (and general) physical phenomena, effects and
mechanisms.

Hence, here we gave an illustrative example which demonstrates how
one can suggest more new effects which is similar in mathematical
descriptions to the familiar phenomena in other areas.
\section{Photonic crystals}
It is well known that the stationary Maxwellian equation for a
time harmonic wave in a medium is exactly analogous to the
stationary Schr\"{o}dinger equation of electrons in a potential
field, namely, the following equation (with the magnetic
permeability $\mu=1$)
\begin{equation}
\nabla^{2}{\bf E}+\omega^{2}\epsilon_{0}\mu_{0}\epsilon{\bf E}=0
\end{equation}
can be rewritten as
\begin{equation}
\nabla^{2}{\bf E}+\omega^{2}\epsilon_{0}\mu_{0}{\bf
E}+\omega^{2}\epsilon_{0}\mu_{0}\chi{\bf E}=0    \label{maxwell}
\end{equation}
with the electric susceptibility $\chi=\epsilon-1$. It is clearly
seen that the form of the equation (\ref{maxwell}) completely
resembles the following stationary Schr\"{o}dinger equation
\begin{equation}
\nabla^{2}\psi+\frac{2m}{\hbar^{2}}E\psi-\frac{2m}{\hbar^{2}}V\psi=0.
\label{Schroding}
\end{equation}

In the solid state physics, a theorem (Bloch's theorem) relating
to the quantum mechanics of crystals, where large numbers of atoms
are held closely together in a lattice, states that the wave
function for an electron in a periodic potential $V({\bf r})$ can
be rewritten as the product of the space harmonic factor $\exp
\left(i{\bf k}\cdot{\bf r}\right)$ and a periodic function. Thus,
Bloch's theorem is interpreted to mean that the wave function for
an electron in a periodic potential field is a plane wave
modulated by a periodic function. It follows that the energy
function of electrons with a lattice-periodicity Hamiltonian is
necessarily multivalued, and separates into branches or bands.
This, therefore, means that in a crystal, electrons are influenced
by a number of adjacent nuclei and the sharply defined levels of
the atoms become bands of allowed energy. Each band, which
represents a large number of allowed quantum states, is separated
from neighboring band by a forbidden band of energies, {\it i.e.},
between the bands are forbidden bands. The energy band theory is
the fundamental approach to crystal and semiconductor physics.

In view of the above discussion, we can inevitably draw our
inspiration from the analogy between (\ref{maxwell}) and
(\ref{Schroding}): if, for example, the electric permittivity (and
hence the electric susceptibility $\chi$) of a certain media is of
the lattice-periodicity structure, then the media will generate a
photonic band gap phenomenon, which would be of interest to many
researchers in materials science, electromagnetism and solid state
physics. Indeed, during the last decades, a kind of material
termed photonic crystals, which is patterned with a periodicity in
dielectric constant and can therefore create a range of forbidden
frequencies called a photonic band gap, focus considerable
attention of a great number of investigators\cite{Yablonovitch}.
Such dielectric structure of crystals offers the possibility of
molding the flow of light inside media. It has many impressive and
striking applications, including the reflecting dielectric,
resonant cavity, waveguide and so on.

The field of photonic crystals is a marriage of solid-state
physics and electromagnetism. Crystal structures are citizens of
solid-state physics, but in photonic crystals the electrons are
replaced by electromagnetic waves. We can see that the concept of
the photonic crystals could be inspired by that of the electronic
crystals.

\section{Left-handed media}
The antiparticle is a subatomic particle that has the same mass as
another particle and equal but opposite values of some other
property or properties. For example, the antiparticle of the
electron is the positron, which has a positive charge equal in
magnitude to the electron's negative charge. Historically, the
existence of antiparticles was predicted from the relativistic
quantum mechanics by Dirac in 1928. It is well known that
Einstein's energy-momentum relation for a particle is of quadratic
form, {\it i.e.}, $E^{2}=p^{2}c^{2}+m_{0}^{2}c^{4}$. From the
purely mathematical point of view, $E$ has two roots, {\it i.e.},
$E_{\pm}=\pm \sqrt{p^{2}c^{2}+m_{0}^{2}c^{4}}$. In the physics of
the 19th century, such a negative root could be regarded as a one
that has no physical meanings and could be discarded without any
hesitation. In quantum mechanics, however, the completely
different things may occur. By solving Dirac's electron wave
equation, one can find that there are solutions corresponding to
the negative root $-\sqrt{p^{2}c^{2}+m_{0}^{2}c^{4}}$ other than
the ones belonging to the positive root
$+\sqrt{p^{2}c^{2}+m_{0}^{2}c^{4}}$. If someone discards the
solutions corresponding to the negative energy root, then the
completeness condition of the solutions of the wave equation no
longer holds. Such a fact shows that the only retained solutions
(corresponding to the positive energy) is not self-consistent in
mathematics, and therefore no functions can be expanded as series
in terms of the solutions corresponding to the positive energy
root only. For this reason, we think the negative-energy solutions
of Dirac's equation should not be discarded. Further analysis
shows that these negative-energy solutions belong to a particle
that has a positive charge and an equal mass of the electron. This
particle is just the positron, the antiparticle of the electron.

The above brief history of the prediction of antiparticle will
unavoidably enlighten us about a lesson: for any physical
phenomena, the mathematical relations of which possess quadratic
forms, we should pay more attention to one of the two roots,
which, at first sight, seems to have no explicit physical
meanings. If we consider it further, it may be possible for us to
find its hidden physical meanings and even to discover another
half world, which had never been familiar to us before. As far as
the negative-energy solutions of Dirac's equation is concerned,
the anti-matter world was discovered by Dirac.

Now let us take into account of the quadratic relation between the
optical refractive index $n$ and the electric permittivity
$\epsilon$ {\it and} magnetic permeability $\mu$, {\it i.e.},
\begin{equation}
n^{2}=\epsilon\mu.
\end{equation}
The solutions has four categories, which are listed as follows
\begin{equation}
n_{1}=+\sqrt{\epsilon\mu}, \quad n_{2}=-\sqrt{\epsilon\mu}, \quad
n_{3}=+\sqrt{(-\epsilon)(-\mu)}, \quad
n_{4}=-\sqrt{(-\epsilon)(-\mu)}.
\end{equation}
With the help of Maxwellian equations, one can discover that only
the cases of $n_{1}=+\sqrt{\epsilon\mu}$ and
$n_{4}=-\sqrt{(-\epsilon)(-\mu)}$ can allow the wave propagation
in media. So, here we will not further consider the solutions
$n_{2}$ and $n_{3}$. If both $\epsilon$ and $\mu$ are positive
numbers, then it follows that there may exist a new kind of
electromagnetic materials, which possess negative electric
permittivity, negative magnetic permeability and hence negative
optical index of refraction!

Now let us turn to the history of research of this new
electromagnetic media: more recently, a kind of artificial
composite metamaterials (the so-called {\it left-handed media})
having a frequency band where the effective permittivity and the
effective permeability are simultaneously negative attracts
considerable attention of many authors both experimentally and
theoretically
\cite{Smith,Klimov,Shelby,Ziolkowski2,Kong,Garcia,Jianqi,Shenlf,Simovski,Lu,He}.
In 1967\footnote{Note that, in the literature, some authors
mentioned the wrong year when Veselago suggested the {\it
left-handed media}. They claimed that Veselago proposed or
introduced the concept of {\it left-handed media} in 1968 or 1964.
On the contrary, the true history is as follows: Veselago's
excellent paper was first published in Russian in July, 1967 [Usp.
Fiz. Nauk {\bf 92}, 517-526 (1967)]. This original paper was
translated into English by W.H. Furry and published again in 1968
in the journal of Sov. Phys. Usp. \cite{Veselago}. Unfortunately,
Furry stated erroneously in his English translation that the
original version of Veselago' work was first published in 1964.},
Veselago first considered this peculiar medium and showed from
Maxwellian equations that such media possessing negative
simultaneously negative $\epsilon $ and $\mu $ exhibit a negative
index of refraction\cite{Veselago}. It follows from the Maxwell's
curl equations that the phase velocity of light wave propagating
inside this medium is pointed opposite to the direction of energy
flow, that is, the Poynting vector and wave vector of
electromagnetic wave would be antiparallel, {\it i.e.}, the vector
{\bf {k}}, the electric field {\bf {E}} and the magnetic field
{\bf {H}} form a left-handed system; thus Veselago referred to
such materials as ``left-handed'' media, and correspondingly, the
ordinary medium in which {\bf {k}}, {\bf {E}} and {\bf {H}} form a
right-handed system may be termed the ``right-handed'' one. Other
authors call this class of materials ``negative-index media
(NIM)'' \cite{Gerardin}, ``double negative media (DNM)''
\cite{Ziolkowski2} and Veselago's media. It is readily verified
that in such media having both $\varepsilon$ and $\mu$ negative,
there exist a number of peculiar electromagnetic and optical
properties, for instance, many dramatically different propagation
characteristics stem from the sign change of the optical
refractive index and phase velocity, including reversal of both
the Doppler shift and Cherenkov radiation, anomalous refraction,
modified spontaneous emission rates, unconventional photon
tunnelling effect, amplification of evanescent wave and even
reversals of radiation pressure to radiation tension
\cite{Klimov}. In experiments, this artificial negative electric
permittivity media may be obtained by using the {\it array of long
metallic wires} (ALMWs) \cite{Pendry2}, which simulates the plasma
behavior at microwave frequencies, and the artificial negative
magnetic permeability media may be built up by using small
resonant metallic particles, {\it e.g.}, the {\it split ring
resonators} (SRRs), with very high magnetic polarizability
\cite{Pendry1,Pendry3,Maslovski}. A combination of the two
structures yields a left-handed medium. Recently, Shelby {\it et
al.} reported their first experimental realization of this
artificial composite medium, the permittivity and permeability of
which have negative real parts \cite{Shelby}. One of the potential
applications of negative refractive index materials is to
fabricate the so-called ``superlenses'' (perfect lenses):
specifically, a slab of such materials may has the power to focus
all Fourier components of a 2D image, even those that do not
propagate in a radiative manner \cite{Pendry2000,Hooft}.

In the paper\cite{cond}, we have demonstrated that the
electromagnetic wave propagation in the negative refractive index
media behaves like that of {\it antiphotons}, which implies that
in certain artificial composite metamaterials (such as the
left-handed media )the complex vector field theory will be a very
convenient theoretical tool for taking into consideration the wave
propagation behavior ({\it e.g.}, scattering, transmission and
refraction).

The investigation of the antiparticle solutions of Dirac equation
means the discovery of another half matter world. Similarly, the
consideration of the negative optical constants ($\epsilon, \mu,
n$) means the discovery of another half material world.
\section{EIT media}
Controlling the phase coherence in ensembles of multilevel atoms
has led to the observation of many striking phenomena in the
propagation of near-resonant light. These phenomena include the
coherent population trapping (CPT), lasing without inversion,
electromagnetically induced transparency (EIT), and anomalously
slow and anomalously fast pulse velocities\cite{Purdy}. Here we
will briefly discuss the coherent population trapping (CPT) and
electromagnetically induced transparency (EIT).

The CPT phenomenon can be observed in a simple three-level atomic
system. The Hamiltonian of such a three-level quantum system in
the interaction picture is given as
\begin{equation}
H_{\rm I}=g_{23}E_{a}\left(|2\rangle\langle 3|+|3\rangle\langle
2|\right)+g_{13}E_{b}\left(|1\rangle\langle 3|+|3\rangle\langle
1|\right),
\end{equation}
where $|1\rangle$ and $|2\rangle$ denote the ground states of this
three-level system, and $|3\rangle$ the excited state. Here the
$E_{a}$ field is tuned to resonance with the transition between
atomic levels $|2\rangle$ and $|3\rangle$, while the $E_{b}$ field
excites the transition between levels $|1\rangle$ and $|3\rangle$.
In order to realize the atomic coherent population trapping, the
excited state $|3\rangle$ should be empty, and the atomic initial
state should be
\begin{equation}
|\Psi(t=0)\rangle=\frac{1}{\sqrt{\left(g_{23}E_{a}\right)^{2}+
\left(g_{13}E_{b}\right)^{2}}}\left(g_{23}E_{a}|1\rangle-g_{13}E_{b}|2\rangle\right).
\label{initialstate}
\end{equation}
It follows from (\ref{initialstate}) that in this state, the
three-level quantum system will not evolve. The physical reason
for this is as follows: since the probability amplitude of level
$|1\rangle$ is $g_{23}E_{a}$, and the interaction strength between
levels $|1\rangle$ and $|3\rangle$ is $g_{13}E_{b}$, the driving
contribution of level $|1\rangle$ to $|3\rangle$, which is the
product of probability amplitude and interaction strength, is
given
\begin{equation}
{\mathcal A}_{1\rightarrow 3}=g_{23}g_{13}E_{a}E_{b}.
\end{equation}
In the meanwhile, the probability amplitude of level $|2\rangle$
is $-g_{13}E_{b}$, and the coupling strength between levels
$|2\rangle$ and $|3\rangle$ is $g_{23}E_{a}$. So, the driving
contribution of level $|2\rangle$ to $|3\rangle$ is
\begin{equation}
{\mathcal A}_{2\rightarrow 3}=-g_{23}g_{13}E_{a}E_{b}.
\end{equation}
Thus, the total driving contributions of the two ground states
$|1\rangle$ and $|2\rangle$ to the excited state $|3\rangle$
vanishes, {\it i.e.},
\begin{equation}
{\mathcal A}_{\rm tot}={\mathcal A}_{1\rightarrow 3}+{\mathcal
A}_{2\rightarrow 3}=0.
\end{equation}
It is apparently seen that the mechanism of atomic CPT is just the
destructive interference. Such a state (\ref{initialstate}), which
can be said to be decoupled to the electromagnetic fields, is
called the dark state (non-coupling state or trapped state). Once
the three-level atom is in the dark state, the transitions from
the ground states to the excited will not occur, and consequently
this atomic ensemble will not absorb the near-resonant lase fields
propagating inside it.

EIT is such a quantum optical phenomenon that if we propagate one
laser beam through a medium and it will get absorbed; but if we
propagate two laser beams instead through the same medium and
neither will be absorbed. Thus the opaque medium is turned into a
transparent one. The physical essence of EIT is just the CPT.
According to the theoretical analysis of multilevel atomic phase
coherence, the requirement of the occurrence of EIT is that the
strength of coupling light is much stronger than that of probe
light\cite{Harris2,Lukin}. This requirement can be interpreted as
follows: it follows that if the laser field $E_{a}$ is much
stronger than $E_{b}$, then the dark state
\begin{equation}
|\Psi(t=0)\rangle\rightarrow |1\rangle,
\end{equation}
which means that under this condition ($E_{a}\gg E_{b}$), one of
the ground levels ({\it i.e.}, $|1\rangle$) can just be thought of
as the dark state. Under this condition, the EIT atomic vapor
allows the probe light to propagate without dissipation through
the medium. From the fully quantum point of view, the interaction
Hamiltonian of the three-level atom interacting with the two
quantized light field is given
\begin{equation}
V=g'_{23}\left(a^{\dagger}|2\rangle\langle 3|+a|3\rangle\langle
2|\right)+g'_{13}\left(b^{\dagger}|1\rangle\langle
3|+b|3\rangle\langle 1|\right),
\end{equation}
where $a^{\dagger}$ ($a$) and $b^{\dagger}$ ($b$) denote the
creation (annihilation) operators of $E_{a}$ and $E_{b}$ fields,
respectively. The effective Hamiltonian of the interaction between
the two ground states is
\begin{equation}
V_{\rm eff1}=Ag'_{13}g'_{23}\left(a^{\dagger}b|2\rangle\langle
1|+b^{\dagger}a|1\rangle\langle 2|\right),
\end{equation}
where $A$ is a certain constant. Thus the effective Hamiltonian of
the four-photon interaction is of the form
\begin{equation}
V_{\rm
eff2}=B\left(g'_{13}g'_{23}\right)^{2}\left[a^{\dagger}b^{\dagger}ba\left(|1\rangle\langle
1|+|2\rangle\langle
2|\right)\right]=\left(g'_{13}g'_{23}\right)^{2}\left(1-|3\rangle\langle
3|\right)a^{\dagger}b^{\dagger}ba
\end{equation}
with $B$ being a certain constant. Here the four-photon
interaction term $a^{\dagger}b^{\dagger}ba$ is similar to the
electron's interaction Hamiltonian in the BCS
(Bardeen-Cooper-Schrieffer) theory of superconductivity. The BCS
theory shows that the current carried in superconductors is not
formed by the individual electrons but by the bound pairs of
electrons, the Cooper pairs. It is seemingly seen that the EIT
phenomenon is truly analogous to the superconductivity. Moreover,
their effective interaction Hamiltonians are almost alike. In EIT,
the two photons of $E_{a}$ and $E_{b}$ fields seem to form a
so-called photonic Cooper pair\cite{Shenzhu}. Here, such photonic
Cooper pairs is intermediated by the two ground states $|1\rangle$
and $|2\rangle$, and the coupling of $|1\rangle$ to $|2\rangle$
results from their interactions with the excited state
$|3\rangle$. We think the above analogy between EIT and
superconductivity is of much interest, and deserves further
investigation.

Now let us consider Cabbibo's theory (1963)\cite{Cabbibo} and the
GIM mechanism (1970) in Weak Interaction. Cabbibo's theory shows
that in the weak interaction, the ${\rm d}$ and ${\rm s}$ quarks
actually behave like the mixed ones ${\rm d}'$ and ${\rm s}'$,
which are the linear combinations
\begin{equation}
{\rm d}'={\rm d}\cos \theta_{\rm c}+{\rm s}\sin \theta_{\rm c},
\quad {\rm s}'=-{\rm d}\sin \theta_{\rm c}+{\rm s}\cos \theta_{\rm
c},
\end{equation}
where $\theta_{\rm c}$ denotes the Cabbibo angle. This means that
the existing form of ${\rm d}$ and ${\rm s}$ quarks in the weak
interaction is in fact the ${\rm d}'$ and ${\rm s}'$ rather than
the ${\rm d}$ and ${\rm s}$ itself. So, the interaction
eigenstates of the above two quarks in the weak interaction is
just the ${\rm d}'$ and ${\rm s}'$ rather than the ${\rm d}$ and
${\rm s}$. Thus the interaction weak current between ${\rm d}'$
and ${\rm u}$ is given
\begin{equation}
\bar{{\rm u}}\gamma_{\mu}\left(1+\gamma_{5}\right){\rm
d}'=\bar{{\rm u}}\gamma_{\mu}\left(1+\gamma_{5}\right)\left({\rm
d}\cos \theta_{\rm c}+{\rm s}\sin \theta_{\rm c}\right).
\end{equation}
The coupling constant of the ${\rm d}\bar{{\rm u}}$ coupling, in
which the strangeness number does not change ($\Delta S=0$), is
$G_{F}$, and the coupling constant of the ${\rm s}\bar{{\rm u}}$
coupling, where the strangeness number changes ($\Delta S=1$), is
$G'_{F}$. Here $G_{F}=G_{\mu}\cos \theta_{\rm c}$,
$G'_{F}=G_{\mu}\sin \theta_{\rm c}$, where $G_{\mu}$ is the
coupling constant of the purely leptonic process (universal
coupling constant in the weak interaction). Thus the Fermi
transition amplitude of ${\rm d}'\rightarrow {\rm u}$ decay is
proportional to
\begin{equation}
\cos \theta_{\rm c}G_{F}+\sin \theta_{\rm c}G'_{F}=G_{\mu},
\end{equation}
while the transition amplitude of ${\rm s}'\rightarrow {\rm u}$
decay is vanishing, {\it i.e.},
\begin{equation}
-\sin \theta_{\rm c}G_{F}+\cos \theta_{\rm
c}G'_{F}=-G_{\mu}\left(\sin \theta_{\rm c}\cos \theta_{\rm c}-\cos
\theta_{\rm c}\sin \theta_{\rm c}\right)=0.
\end{equation}
In other words, the ${\rm s}'$ quark can be said to be steady for
the case of decay into ${\rm u}$. This, therefore, means that in
the weak interaction, it is more reasonable and convenient for
$({\rm d}', {\rm s}')$ to represent the interaction eigenstate
than for $({\rm d}, {\rm s})$.

In 1970's, in the experiments physicists found that all the
weak-interaction neutral current processes satisfy the selection
rule of $\Delta S=0$, while the neutral current process of $\Delta
S=1$ could rarely be observed. In accordance with the ${\rm u}{\rm
d}{\rm s}$ three-quark theory, however, there should exist the
neutral current process of $\Delta S=1$. For example, the
weak-interaction matrix element of neutral currents ${\rm
u}\bar{{\rm u}}\rightarrow z^{0}$ and ${\rm d}'\bar{{\rm
d}'}\rightarrow z^{0}$ takes the form
\begin{equation}
{\rm u}\bar{{\rm u}}+\left({\rm d}\bar{{\rm d}}\cos^{2}
\theta_{\rm c}+{\rm s}\bar{{\rm s}}\sin^{2} \theta_{\rm
c}\right)+\left({\rm s}\bar{{\rm d}}+\bar{{\rm s}}{\rm
d}\right)\sin \theta_{\rm c}\cos \theta_{\rm c},
\end{equation}
where ${\rm u}\bar{{\rm u}}+\left({\rm d}\bar{{\rm d}}\cos^{2}
\theta_{\rm c}+{\rm s}\bar{{\rm s}}\sin^{2} \theta_{\rm c}\right)$
is the contribution of the $\Delta S=0$ process, and $\left({\rm
s}\bar{{\rm d}}+\bar{{\rm s}}{\rm d}\right)\sin \theta_{\rm c}\cos
\theta_{\rm c}$ results from the $\Delta S=1$ process. Note that
in theory there truly exists the neutral current of $\Delta S=1$.
How can we explain this conflict between theory and experiments?
In 1970, Glashow {\it et al.} proposed a mechanism (which is now
known as GIM mechanism) and supposed that there exists a new quark
(${\rm c}$) that had not been discovered in Nature. In the GIM
mechanism, $({\rm u}, {\rm d}')$ and $({\rm c}, {\rm s}')$ form
the weak-interaction doublets, respectively, {\it i.e.},
\begin{equation}
\left( {\begin{array}{*{20}c}
   {{\rm u}}  \\
   {{\rm
d}'}  \\
\end{array}} \right)=\left( {\begin{array}{*{20}c}
   {{\rm u}}  \\
   {{\rm d}\cos \theta_{\rm c}+{\rm s}\sin \theta_{\rm c}}  \\
\end{array}} \right),   \quad
\left( {\begin{array}{*{20}c}
   {{\rm c}}  \\
   {{\rm
s}'}  \\
\end{array}} \right)=\left( {\begin{array}{*{20}c}
   {{\rm c}}  \\
   {-{\rm d}\sin \theta_{\rm c}+{\rm s}\cos \theta_{\rm
c}}  \\
\end{array}} \right).
\end{equation}
According to the GIM mechanism, in addition to the neutral
currents ${\rm u}\bar{{\rm u}}\rightarrow z^{0}$ and ${\rm
d}'\bar{{\rm d}'}\rightarrow z^{0}$, there exist another two
neutral currents ${\rm c}\bar{{\rm c}}\rightarrow z^{0}$ and ${\rm
s}'\bar{{\rm s}'}\rightarrow z^{0}$. Thus, the weak-interaction
matrix element of the four neutral currents ${\rm u}\bar{{\rm
u}}\rightarrow z^{0}$, ${\rm d}'\bar{{\rm d}'}\rightarrow z^{0}$,
${\rm c}\bar{{\rm c}}\rightarrow z^{0}$ and ${\rm s}'\bar{{\rm
s}'}\rightarrow z^{0}$ can be written in the form
\begin{equation}
{\rm u}\bar{{\rm u}}+{\rm c}\bar{{\rm c}}+\left[\left({\rm
d}\bar{{\rm d}}+{\rm s}\bar{{\rm s}}\right)\cos^{2} \theta_{\rm
c}+\left({\rm s}\bar{{\rm s}}+{\rm d}\bar{{\rm d}}\right)\sin^{2}
\theta_{\rm c}\right]+\left({\rm s}\bar{{\rm d}}+\bar{{\rm s}}{\rm
d}-\bar{{\rm s}}{\rm d}-{\rm s}\bar{{\rm d}}\right)\sin
\theta_{\rm c}\cos \theta_{\rm c},
\end{equation}
where ${\rm u}\bar{{\rm u}}+{\rm c}\bar{{\rm c}}+\left[\left({\rm
d}\bar{{\rm d}}+{\rm s}\bar{{\rm s}}\right)\cos^{2} \theta_{\rm
c}+\left({\rm s}\bar{{\rm s}}+{\rm d}\bar{{\rm d}}\right)\sin^{2}
\theta_{\rm c}\right]$ is the contribution of the neutral current
of $\Delta S=0$, and the contribution of $\Delta S=1$ is
\begin{equation}
\left({\rm s}\bar{{\rm d}}+\bar{{\rm s}}{\rm d}-\bar{{\rm s}}{\rm
d}-{\rm s}\bar{{\rm d}}\right)\sin \theta_{\rm c}\cos \theta_{\rm
c}=0,
\end{equation}
which demonstrates that the neutral current of $\Delta S=1$ is
automatically eliminated and will therefore not exist in the weak
interaction. In a word, the introduction of ${\rm c}$ quark leads
to the destructive interference of the neutral current of $\Delta
S=1$ among the weak-interaction doublets $({\rm u}, {\rm d}')$ and
$({\rm c}, {\rm s}')$. Thus, the GIM mechanism resolves
successfully the above conflict between the ${\rm u}{\rm d}{\rm
s}$ three-quark theory and the experiments.

We think the destructive interference is one of the common
features in EIT (CPT) and Cabbibo's theory {\it and} GIM
mechanism.
\\ \\
\textbf{Acknowledgements}  This work was supported by the National
Natural Science Foundation of China under Project No. $90101024$
and $60378037$.


\begin{references}

\bibitem{Yablonovitch} E. Yablonovitch, Phys. Rev. Lett.
\textbf{58}, 2059 (1987); E. Yablonovitch and T.J. Gmitter, Phys.
Rev. Lett. \textbf{63}, 1950 (1989); P. Villeneuve and M. Piche,
Phys. Rev. B \textbf{46}, 4969 (1992).


\bibitem{Smith} D.R. Smith, W.J. Padilla, D.C. Vier et
al., Phys. Rev. Lett. \textbf{84}, 4184 (2000).

\bibitem{Klimov} V.V. Klimov, Opt. Comm. \textbf{211}, 183
(2002).

\bibitem{Shelby} R.A. Shelby, D.R. Smith, and S. Schultz,
Science \textbf{292}, 77 (2001).

\bibitem{Ziolkowski2} R.W. Ziolkowski, Phys. Rev. E \textbf{64},
056625 (2001).

\bibitem{Kong} J.A. Kong, B.L. Wu, and Y. Zhang, Appl. Phys. Lett.
\textbf{80}, 2084 (2002).

\bibitem{Garcia} N. Garcia and M. Nieto-Vesperinas, Opt. Lett. \textbf{27}, 885 (2002).

\bibitem{Jianqi} J.Q. Shen, Phys. Scr. \textbf{68}, 87 (2003).

\bibitem{Shenlf} L.F. Shen and S.L. He, Phys. Lett. A \textbf{309}, 298 (2003).

\bibitem{Simovski}  C.R. Simovski and S.L. He, Phys. Lett. A \textbf{311}, 254 (2003).

 \bibitem{Lu}  J Lu and S.L. He, Microwwave Opt. Technol. Lett. \textbf{37}, 292 (2003).

 \bibitem{He}  C.R. Simovski, P.A. Belov, and S.L. He, IEEE Trans. Antennas Propagat. (special issue on metamaterials) 1 (2003).

\bibitem{Veselago} V.G. Veselago, Sov. Phys. Usp. \textbf{10}, 509
(1968).

\bibitem{Gerardin} J. Gerardin and A. Lakhtakia, Phys. Lett. A \textbf{301}, 377 (2002).

\bibitem{Pendry2} J.B. Pendry, A.J. Holden, D.J. Robbins, and
W.J. Stewart, J. Phys. Condens. Matter \textbf{10}, 4785 (1998).

\bibitem{Pendry1} J.B. Pendry, A.J. Holden, W.J. Stewart, and
I. Youngs, Phys. Rev. Lett. \textbf{76}, 4773 (1996).

\bibitem{Pendry3} J.B. Pendry, A.J. Holden, D.J. Robbins, and
W. . Stewart, IEEE Trans. Microwave Theory Tech. \textbf{47}, 2075
(1999).

\bibitem{Maslovski} S.I. Maslovski, S.A. Tretyakov, and P.A.
Belov, Inc. Microwave Opt. Tech. Lett. \textbf{35}, 47 (2001).

\bibitem{Pendry2000} J.B. Pendry, Phys. Rev. Lett. \textbf{85}, 3966 (2000).

\bibitem{Hooft}   G.W. t' Hooft, Phys. Rev. Lett. \textbf{87}, 249701 (2001).

\bibitem{cond} J.Q. Shen, arXiv: cond-mat/0308349 (2003).

\bibitem{Purdy} T. Purdy and M. Ligare, arXiv: quant-ph/0204173
(2002).

\bibitem{Harris2}  S.E. Harris, Phys. Today {\bf 50}(7), 36
(1997).

\bibitem{Lukin} M. Lukin, S. Yellin, A. Zibrov, and M. Scully,
Laser Physics {\bf 9}, 759 (1999).

\bibitem{Shenzhu} J.Q. Shen, H.Y. Zhu, and H.L. Zhu, Laser \&
Infrared (in Chinese) {\bf 32}, 315 (2002).

\bibitem{Cabbibo} N. Cabbibo, Phys. Lett. {\bf 10}, 531 (1963).

\end{references}
\end{document}